\documentclass[twocolumn, superscriptaddress, prb, footinbib]{revtex4}

\usepackage{graphicx}
\usepackage{amsmath}
\usepackage{textcomp}
\usepackage{hyperref}

\begin{document}

\title{The intercalation phase diagram of Mg in V$_2$O$_5$ from first principles}

\author{Gopalakrishnan Sai Gautam} \email{gautam91@mit.edu}
\affiliation{
Department of Materials Science and Engineering, Massachusetts
Institute of Technology, Cambridge, MA 02139, USA}

\author{Pieremanuele Canepa}\affiliation{
Department of Materials Science and Engineering, Massachusetts
Institute of Technology, Cambridge, MA 02139, USA}

\author{Aziz Abdellahi}\affiliation{
Department of Materials Science and Engineering, Massachusetts
Institute of Technology, Cambridge, MA 02139, USA}

\author{Alexander Urban}\affiliation{
Department of Materials Science and Engineering, Massachusetts
Institute of Technology, Cambridge, MA 02139, USA}

\author{Rahul Malik}\affiliation{
Department of Materials Science and Engineering, Massachusetts
Institute of Technology, Cambridge, MA 02139, USA}

\author{Gerbrand Ceder} \email{gceder@mit.edu}
\affiliation{
Department of Materials Science and Engineering, Massachusetts
Institute of Technology, Cambridge, MA 02139, USA}


\begin{abstract}
We have investigated Mg intercalation into orthorhombic V$_2$O$_5$, one of only three cathodes known to reversibly intercalate Mg ions. By calculating the ground state Mg$_x$V$_2$O$_5$ configurations and by developing a cluster expansion for the configurational disorder in $\delta$-V$_2$O$_5$, a full temperature-composition phase diagram is derived. Our calculations indicate an equilibrium phase separating behavior between fully demagnesiated $\alpha$-V$_2$O$_5$ and fully magnesiated $\delta$-V$_2$O$_5$, but also motivate the existence of potentially metastable solid solution transformation paths in both phases. We find significantly better mobility for Mg in the $\delta$ polymorph suggesting that better performance can be achieved by cycling Mg in the $\delta$ phase.
\end{abstract}

\maketitle
\section{Introduction}
\label{sec:intro}
A multi-valent (MV) battery chemistry, which pairs a non-dendrite forming Mg metal anode with a high voltage ($\sim$ 3~V) intercalation cathode offers a potentially safe and inexpensive high energy density storage system with the potential to outperform current Li-ion technology.\cite{Noorden2014} A change in chemistry leads to new challenges, however, one being the design of a cathode that can reversibly intercalate Mg at a high enough voltage. Orthorhombic V$_2$O$_5$ is one such material that offers exciting prospects of being a reversible intercalating cathode for Mg batteries.\cite{Shterenberg2014,Aurbach2000a,Yoo2013} The theoretical energy density of a cathode based on Mg intercalation into V$_2$O$_5$ is $\sim$~660~Wh/kg,\cite{Jain2013} which approaches the practical energy densities of current commercial Li-ion chemistries ($\sim$~700~Wh/kg for LiCoO$_2$\cite{Whittingham2004}), but the major benefit of switching to a MV chemistry is the gain in volumetric energy density arising from the usage of a metallic anode ($\sim$~3833~mAh/cm$^3$ for Mg\cite{Shterenberg2014} compared to $\sim$~800~mAh/cm$^3$ for Li insertion into graphite.\cite{Jain2013})

The orthorhombic V$_2$O$_5$ structure has been well characterized due to its interesting spin ladder characteristics and widely known Li intercalation properties, with a reversible capacity of $\sim$ 130~mAh/g and voltage of $\sim$ 3.3~V vs.\ Li metal.\cite{Bachmann1961,Enjalbert1986,Millet1998b,Korotin2000,Whittingham1976,Galy1965,Amatucci2001a} Consequently, Li intercalation into V$_2$O$_5$ has been the subject of several experimental\cite{Delmas1994,Murphy1979,Dickens1981,Wiesener1987,Wu2005} and theoretical\cite{Braithwaite1999,Rocquefelte2003,Scanlon2008} studies. Li-V$_2$O$_5$ undergoes several first-order phase transformations during intercalation, such as the $\alpha \rightarrow \epsilon$ and $\epsilon \rightarrow \delta$ between x$_\text{Li}$~=~0 and x$_\text{Li} = 1$, the irreversible $\delta \rightarrow \gamma$ transition at x$_\text{Li} > 1$, and another irreversible $\gamma \rightarrow \omega$ transition at x$_\text{Li} > 2$.\cite{Delmas1994} Several authors have investigated Mg-insertion into V$_2$O$_5$\cite{Amatucci2001a,Ramos1987,Gregory1990,Imhof1999,Gershinsky2013} and to date, V$_2$O$_5$ is one of only three cathode materials to have shown reversible intercalation of Mg, the other two being the chevrel Mo$_3$S$_4$\cite{Aurbach2000a} and layered MoO$_3$.\cite{Gershinsky2013}

While Li-ion has been investigated extensively for the past $\sim$ 25 years, there are significantly fewer studies, theoretical or otherwise, of Mg intercalation hosts in the literature. Pereira-Ramos \emph{et al.}\cite{Ramos1987} showed electrochemical intercalation of Mg into V$_2$O$_5$ (at 150~{\textdegree}C and 100~$\mu$A/cm$^2$ current density), and Gregory \emph{et al.}\cite{Gregory1990} have reported chemical insertion of Mg up to Mg$_{0.66}$V$_2$O$_5$. Novak \emph{et al.}\cite{Novak1993} demonstrated reversible electrochemical insertion of Mg in V$_2$O$_5$ at room temperature while also demonstrating superior capacities ($\sim$ 170~mAh/g) using an acetonitrile (AN) electrolyte containing water as opposed to dry AN. Yu \emph{et al.}\cite{Yu2004} showed similar improvements in capacity ($\sim$158.6~mAh/g) using a H$_2$O $+$ Polycarbonate (PC) system compared to dry PC. Electrochemical insertion of Mg into V$_2$O$_5$ nanopowders and thin films using activated carbon as the counter electrode was shown by Amatucci \emph{et al.}\cite{Amatucci2001a} and Gershinsky \emph{et al.},\cite{Gershinsky2013} respectively, and insertion into V$_2$O$_5$ single crystals was reported by Shklover \emph{et al.}\cite{Shklover1996}
 
Thus far, all reported experimental attempts have begun in the charged state and succeeded in reversibly inserting only about half a Mg (x$_\text{Mg} \sim 0.5$) per formula unit of V$_2$O$_5$, in contrast to Li-V$_2$O$_5$ where up to x$_\text{Li} \sim 3$ has been inserted per V$_2$O$_5$.\cite{Delmas1994,Ramos1987,Yu2004,Shklover1996} When the grain size of V$_2$O$_5$ is reduced, e.g., nano powders and thin films, insertion levels can reach x$_\text{Mg} \sim 0.6$.\cite{Amatucci2001a,Gershinsky2013} In addition, in cells where a Mg metal anode was used rapid capacity fade was reported upon cycling.\cite{Novak1993,Yu2004} Unlike Li intercalation systems, anode passivation by the electrolytes is a major issue for Mg batteries using a Mg metal anode.\cite{Yu2004} Out of the two experiments that have not reported significant capacity fade so far,\cite{Amatucci2001a,Gershinsky2013} the work done by Gershinsky \emph{et al.} is particularly useful to benchmark theoretical models as the Mg insertion was done at extremely low rates (0.5~$\mu$A/cm$^2$), and therefore corresponds most to equilibrium conditions.

Previous theoretical studies of the Mg-V$_2$O$_5$ system have benchmarked structural parameters, average voltages and the electronic properties of layered V$_2$O$_5$ upon Mg insertion.\cite{Wang2013,Carrasco2014,Zhou2014} Wang \emph{et al.}\cite{Wang2013} showed an increase in the Mg binding energy and Li mobility in single-layered V$_2$O$_5$ compared to bulk V$_2$O$_5$. Carrasco\cite{Carrasco2014} found that while incorporating van der Waals dispersion corrections in the calculations improved the agreement of the  lattice parameters with experiments, it led to an overestimation of the voltage. Zhou \emph{et al.}\cite{Zhou2014} calculated the band structures, average voltages, Mg migration barriers, and the $\alpha \rightarrow \delta$ phase transformation barrier in Mg-V$_2$O$_5$. While reporting higher computed average voltage for Mg-V$_2$O$_5$ compared to the Li-V$_2$O$_5$ system (in apparent disagreement with experiments\cite{Delmas1994,Gershinsky2013}), the authors explained the slow diffusion of Mg in V$_2$O$_5$ by predicting a facile $\alpha \rightarrow \delta$ transition coupled with an estimated lower Mg mobility in $\delta$ than $\alpha$.\cite{Zhou2014}

In the present work, we have explored in detail the physics of room temperature Mg intercalation in orthorhombic V$_2$O$_5$ using first-principles calculations. Compared to Li, Mg insertion is accompanied by twice the number of electrons, which means that the properties of the Mg intercalation system will be largely dictated by how the additional electron localizes on the nearby V atoms. To study the combined effects not only of inserting a different ion but also a different number of electrons on the equilibrium phase behavior, we calculate the Mg-V$_2$O$_5$ intercalation phase diagram using the Cluster expansion-Monte Carlo approach. A similar approach has been previously used to study Li-intercalation systems\cite{Ceder1999,Zhou2006} and can be derived formally through systematic coarse graining of the partition function.\cite{Ceder1993} Our calculations focus particularly on Mg intercalation into the $\alpha$ and $\delta$ polymorphs of V$_2$O$_5$, evaluating their respective ground state hulls, subsequent voltage curves and activation barriers for Mg diffusion. We have also constructed the temperature-composition phase diagram for Mg in the $\delta$ polymorph. 

\section{Polymorphs of V$_2$O$_5$}
\label{sec:structure}

\begin{figure*}[]
\includegraphics[scale=0.4]{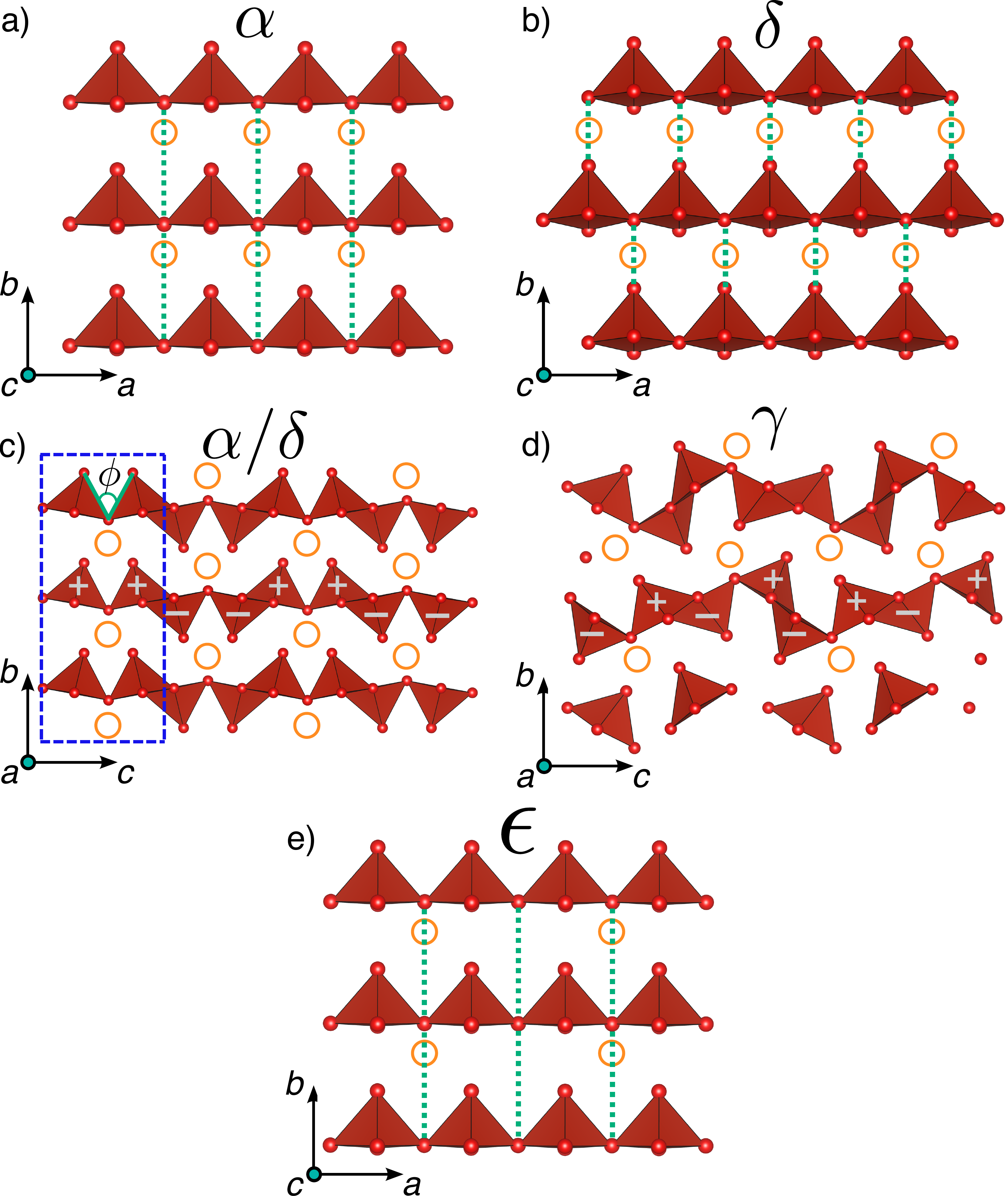}
\caption{
\label{fig:1} (Color online) 
(a) $\alpha$ and (b) $\delta$ polymorphs of orthorhombic V$_2$O$_5$ are shown along the \emph{c}-axis (shown to a depth of \emph{c}/2 for viewing clarity) and along the (c) \emph{a}-axis, which compared to the (d) $\gamma$ polymorph has a different orientation of VO$_5$ pyramids as denoted by `$+$' and `$-$' signs along the \emph{c}-axis. Hollow orange circles correspond to the intercalation sites, the green dotted lines show the differences in layer stacking and the dashed blue rectangle in (c) indicates a distance of \emph{c}/2. (e) illustrates the $\epsilon$ phase corresponding to a specific ordering of Mg atoms in $\alpha$-V$_2$O$_5$  at half magnesiation, where alternate intercalant sites are occupied in the \emph{a} axis as indicated by the orange circles. The schematics here correspond to `supercells' of the respective polymorph unit cells.}
\end{figure*}

The V$_2$O$_5$ structure consists of layers of VO$_5$ pyramids, each of which have 4 V$-$O bonds that form the base of the pyramid and one V$=$O (Vanadyl) bond that forms the apex. Each layer consists of alternate corner and edge sharing pyramids, with an offset in the \emph{a}-axis between the edge-sharing pyramids. The different polymorphs of V$_2$O$_5$ observed experimentally are illustrated in Figure~\ref{fig:1},\cite{Delmas1994} with the $\alpha$ (space group \emph{Pmmn}), $\delta$ (\emph{Cmcm}) and $\gamma$ (\emph{Pnma}) polymorphs all having orthorhombic symmetry. The notation, specific to this work, is \emph{a} being the shortest axis of the lattice (3.56~$\textrm{\AA}$ for $\alpha$; 3.69~$\textrm{\AA}$ for $\delta$), \emph{b} being the axis perpendicular to the layers indicative of the layer spacing (4.37~$\textrm{\AA}$; 9.97~$\textrm{\AA}$), and \emph{c} being the longest axis (11.51~$\textrm{\AA}$; 11.02~$\textrm{\AA}$). Pure V$_2$O$_5$ crystallizes in the $\alpha$ phase at 298~K and remains stable at higher temperatures,\cite{Delmas1994} while the fully magnesiated phase (MgV$_2$O$_5$) has been found to form in the structure of the $\delta$ polymorph.\cite{Bouloux1976} For simpler visualization, a single slice of the $\alpha$ and $\delta$ polymorphs, corresponding to a depth of \emph{c}/2 (illustrated by the dashed blue rectangle in Figure~\ref{fig:1}c) is shown in Figure~\ref{fig:1}a and Figure~\ref{fig:1}b respectively. The $\alpha$ and $\delta$ polymorphs are very similar when viewed along the \emph{a}-axis or the \emph{b-c} plane (Figure~\ref{fig:1}c). 

The main difference between the $\delta$ phase and the $\alpha$ phase is a translation of alternating V$_2$O$_5$ layers in the \emph{a}-direction by `\emph{a}/2' which doubles the `\emph{b}' lattice parameter (as well as the unit cell) of the $\delta$ phase. The Mg sites in both $\alpha$ and $\delta$ are situated near the middle of the VO$_5$ pyramids (along \emph{a}) and between the 2 layers (along \emph{b}), as illustrated by the orange circles in Figure~\ref{fig:1}. As a result of shifting of layers between the $\alpha$ and $\delta$ phases, the anion coordination environment of the Mg sites also changes. Considering a Mg$-$O bond length cutoff of 2.5~$\textrm{\AA}$, the Mg in the $\alpha$ phase is 8-fold coordinated (4 nearest neighbor O atoms and 4 next nearest neighbors, 4$+$4) whereas the Mg in the $\delta$ phase is 6-fold coordinated (4$+$2). In this work, the $\epsilon$ phase is a specific ordering of Mg atoms on the $\alpha$-V$_2$O$_5$ host at half magnesiation, as shown in Figure~\ref{fig:1}e. This intercalant ordering is observed in the Li-V$_2$O$_5$ system,\cite{Delmas1994} and has intercalant ions at alternate sites along the \emph{a} axis, as illustrated by the absence of Mg sites in Figure~\ref{fig:1}e.\cite{Rocquefelte2003,Cocciantelli1991} The VO$_5$ pyramids in the $\alpha$ and $\delta$ phases `pucker' upon Li intercalation as observed experimentally by Cava \emph{et al.}\cite{Cava1986} For the sake of simplicity we define puckering here as the angle `$\phi$', as shown in Figure~\ref{fig:1}c. As the pyramids pucker with intercalation, the angle `$\phi$' decreases. 

In the Li-V$_2$O$_5$ system, at x$_\text{Li} > 1$, the host structure undergoes an irreversible phase transformation to form the $\gamma$ phase, in which the VO$_5$ pyramids adopt a different orientation compared to $\alpha$ and $\delta$, as seen in Figure~\ref{fig:1}c and~\ref{fig:1}d.\cite{Delmas1994} In the $\gamma$ phase, the VO$_5$ pyramids along the \emph{c}-direction alternate between up and down (denoted by `$+$' and `$-$' in Figure~\ref{fig:1}); whereas, in $\alpha$ and $\delta$, the sequence goes as `up-up-down-down'. The $\gamma$ phase has not yet been reported in the Mg-V$_2$O$_5$ system and hence will not be further discussed in this paper. 

\section{Methodology}
\label{sec:methods}
To compute the ground state hull and the average open circuit voltage curves we use Density Functional Theory (DFT) as implemented in VASP with the Perdew-Burke-Ernzerhof (PBE) exchange-correlation functional.\cite{Kohn1965,Kresse1993,Kresse1996,Perdew1996} The Projector Augmented Wave theory\cite{Kresse1999} together with a well converged energy cutoff of 520~eV is used to describe the wave functions, which are sampled on a $\Gamma$-centered 4$\times$4$\times$4 \emph{k}-point mesh. In order to remove the spurious self-interaction of the vanadium \emph{d}-electrons, a Hubbard~\emph{U} correction of 3.1~eV is added to the Generalized Gradient Approximation (GGA) Hamiltonian (GGA+\emph{U})\cite{Anisimov1991,Zhou2004_U} as fitted by Jain \emph{et al.}\cite{Jain2011} All Mg-V$_2$O$_5$ structures are fully relaxed within 0.25~meV/f.u.
 
To obtain the temperature-composition phase diagram, Grand-canonical Monte Carlo (GMC) simulations are performed on a cluster expansion (CE) Hamiltonian. The CE is a parameterization of the total energy with respect to the occupancy of a predefined topology of sites, which in this case are the possible Mg insertion sites.\cite{Ceder1993,Sanchez1984,Anton2000_Sol} In practice the CE is written as a truncated summation of the Effective Cluster Interactions (ECIs) of the pair, triplet, quadruplet and higher order terms as given in Equation~\ref{eq:ce}.

\begin{equation}
\label{eq:ce}
E(\sigma) = \sum\limits_{\alpha} m_\alpha V_\alpha \langle \prod\limits_{i\in\beta} \sigma_i \rangle
\end{equation}

where the energy, $E$ of a given configuration of Mg ions $\sigma$ is obtained as a summation over all symmetrically distinct clusters $\alpha$. Each term in the sum is a product of the multiplicity $m$, the effective cluster interaction (ECI) $V$ for a given $\alpha$, and the occupation variable $\sigma_i$ averaged over all clusters $\beta$ that are symmetrically equivalent to $\alpha$ in the primitive cell of the given lattice. In this work, the CE is performed on the Mg sub-lattice and the various configurations correspond to the arrangement of Mg ($\sigma_i = 1$) and Vacancies (Va; $\sigma_i = -1$) on the available Mg sites. The Pymatgen library is used to generate the various Mg-Va arrangements to be calculated with DFT.\cite{Ong2013,Hart2008,Hart2009,Hart2012} The CE is built on the DFT formation energy of 97 distinct Mg-Va configurations using the compressive sensing paradigm and optimized through the split-Bregman algorithm.\cite{Nelson2013,Goldstein2009} The root mean square error (RMSE) and the weighted cross-validation (WCV) score are used to judge the quality and the predictive ability of the fit, respectively.\cite{Axel2002}

The high temperature phase diagram is then obtained with GMC calculations on supercells  containing at least 1728 Mg/Va sites (equivalent to a 12$\times$6$\times$6 supercell of the conventional unit cell) and for a minimum of 100,000 equilibration steps followed by 200,000 sampling steps.\cite{VanderVen2010}  Monte Carlo scans are done on a range of chemical potentials at different temperatures, and phase transitions are detected by discontinuities in Mg concentration and energies. In order to remove numerical hysteresis from the Monte Carlo simulations, particularly at low temperatures, free energy integration is performed\cite{Hinuma2008} with the fully magnesiated and fully demagnesiated phases as reference states.

Finally, the activation barriers associated with Mg diffusion in V$_2$O$_5$ are calculated with DFT using the Nudged Elastic Band method (NEB)\cite{Sheppard2008} and forces converged within 100~meV/$\textrm{\AA}$. A minimum distance of 9~$\textrm{\AA}$ is introduced between the diffusing species and nine distinct images are used to capture the diffusion trajectory. As previously indicated by Liu \emph{et al.},\cite{Liu2014a} the convergence of GGA+\emph{U} NEB calculations is problematic, and hence standard GGA is used to compute the Mg diffusion barriers.

\section{Results}
\label{sec:results}
\subsection{Mg-V$_2$O$_5$ Ground State Hull}
\label{subsec:hull}
\begin{figure}[h]
\includegraphics[width=\columnwidth]{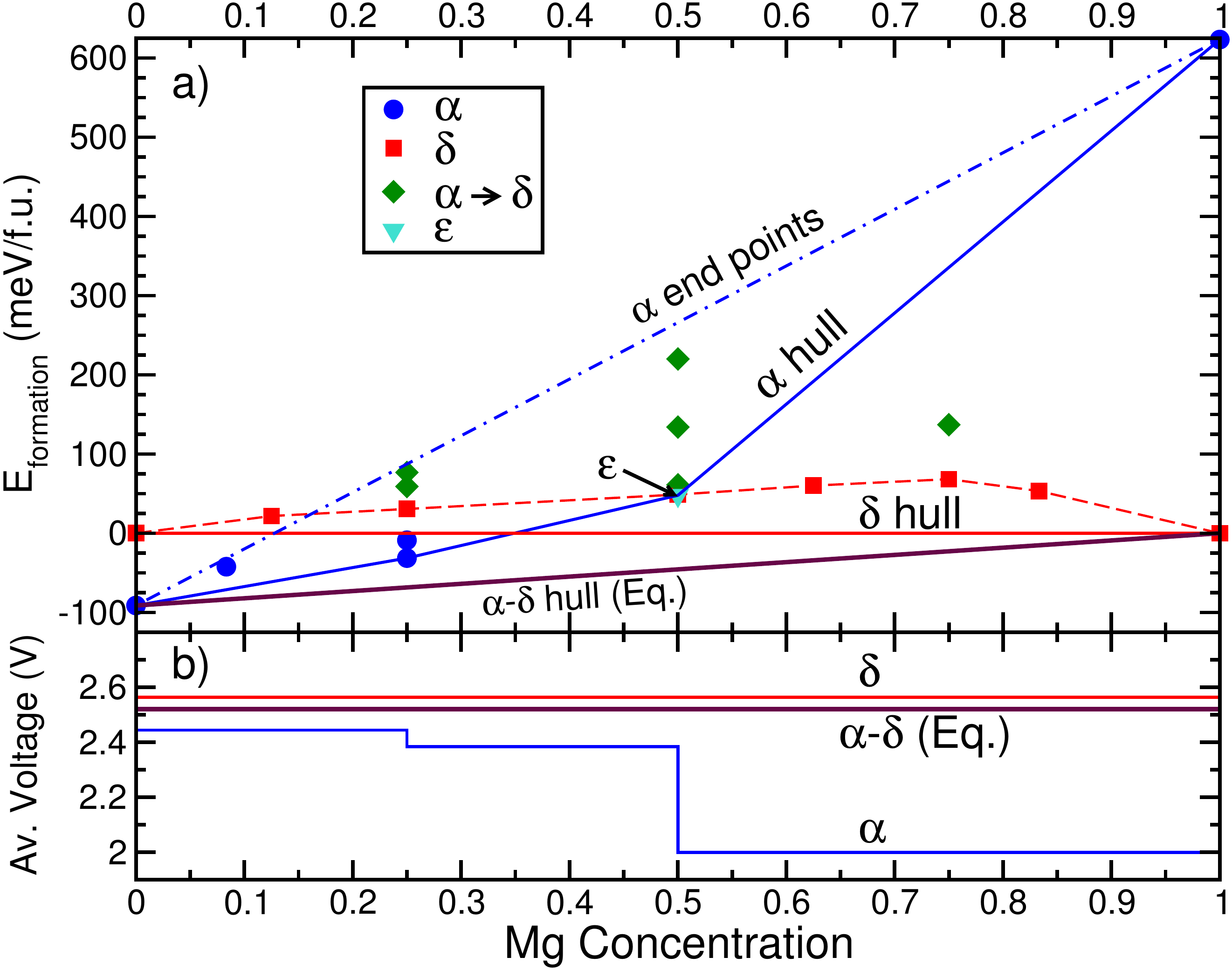}
\caption{
\label{fig:2} (Color online) 
(a) The ground-state hull of Mg in V$_2$O$_5$ considering both $\alpha$ and $\delta$ phases. The formation energy per formula unit has been plotted with respect to Mg concentration. (b) The average voltage curves at 0~K for the $\alpha$ and $\delta$ phases with respect to pure Mg metal, obtained from the respective hulls are plotted against the Mg concentration.}
\end{figure}

Figure~\ref{fig:2} shows the ground state hull and average voltage curves as a function of Mg concentration in V$_2$O$_5$ as computed by DFT. The solid blue and red lines in Figure~\ref{fig:2}a indicate the ground state hulls of the $\alpha$ and $\delta$ polymorphs respectively. All formation energies are referenced to the fully magnesiated and fully demagnesiated end points of the $\delta$-phase. The overall equilibrium behavior of the system is that of phase separation between unintercalated $\alpha$-V$_2$O$_5$ and fully intercalated $\delta$-Mg$_1$V$_2$O$_5$ as indicated by the solid maroon line. As can be observed, the $\alpha$ phase is stable compared to the $\delta$ phase at low Mg concentrations up to x$_\text{Mg} \sim 0.35$ where the $\alpha$ and $\delta$ hulls intersect, and the $\delta$ phase is stable at higher Mg concentrations. In Figure~\ref{fig:2}a, the dash-dotted blue line indicates the end members of the $\alpha$ hull (pure $\alpha$-V$_2$O$_5$ and $\alpha$-Mg$_1$V$_2$O$_5$), and the dashed red line the lowest energy configurations computed at intermediate Mg concentrations for the $\delta$ phase. 

The $\alpha$-hull represents the energy trajectory for metastable Mg insertion into $\alpha$-V$_2$O$_5$ (i.e., without transformation of the host to $\delta$), and it displays a convex shape with ground state configurations at Mg concentrations of 0.25 and 0.5. The most stable configuration at x$_\text{Mg} = 0.5$ in the $\alpha$ hull is the $\epsilon$ phase. In contrast, there are no metastable Mg orderings in the $\delta$ phase implying that in the $\delta$-phase host the Mg ions will want to phase separate into MgV$_2$O$_5$ and V$_2$O$_5$ domains. Some Mg configurations when initialized in the $\alpha$ phase relax to the $\delta$ phase as indicated by the green diamond points on Figure~\ref{fig:2}a. These structures undergo a shear-like transformation from $\alpha$ to $\delta$, which involves V$_2$O$_5$ layers sliding along the \emph{a}-direction. This mechanical instability phenomenon has been observed in our calculations both at low Mg concentrations (x$_\text{Mg} = 0.25$) and at high Mg concentrations (x$_\text{Mg} = 0.75$), but never at very low Mg concentrations (x$_\text{Mg} = 0.08$). 

The Mg insertion voltage will depend on which of the possible stable or metastable paths the system follows and the voltage for several possible scenarios is shown in Figure~\ref{fig:2}b. The equilibrium voltage curve is a single plateau at 2.52~V vs. Mg metal, consistent with phase separating behavior between $\alpha$-V$_2$O$_5$ and $\delta$-Mg$_1$V$_2$O$_5$. The voltage for the metastable insertion in the $\alpha$ host averages $\sim$ 2.27~V vs. Mg metal for $0 <$~x$_\text{Mg} < 1$ and exhibits a steep potential drop of $\sim$ 400~mV at x$_\text{Mg} = 0.5$, corresponding to the $\epsilon$ ordering. Metastable Mg insertion in $\delta$ occurs on a single plateau at 2.56~V vs. Mg metal, consistent with phase separation between $\delta$-Mg$_0$V$_2$O$_5$ and fully intercalated $\delta$-Mg$_1$V$_2$O$_5$. The average voltage of the $\alpha$ phase best agrees with the experimental average voltage of $\sim$ 2.3~V.\cite{Amatucci2001a,Gershinsky2013}

\subsection{Puckering and Layer spacing}
\label{subsec:layer}
The VO$_5$ pyramids in both $\alpha$ and $\delta$-V$_2$O$_5$ pucker upon Mg intercalation, quantified by the angle $\phi$ shown in Figure~\ref{fig:1}c. We find that $\phi$ decreases (corresponding to increased puckering) with increasing Mg concentration, resulting in the formation of ripples in the layers. Current calculations show a decrease from $\phi \sim$~76{\textdegree} at x$_\text{Mg} = 0$ (which corresponds to flat layers) to $\phi \sim$~56{\textdegree} at x$_\text{Mg} = 1$ in the $\alpha$ phase and a decrease from $\phi \sim$~68{\textdegree} at x$_\text{Mg} = 0$ to $\phi \sim$~54{\textdegree} at x$_\text{Mg} = 1$ in $\delta$-V$_2$O$_5$. 

\begin{figure}[h]
\includegraphics[width=\columnwidth]{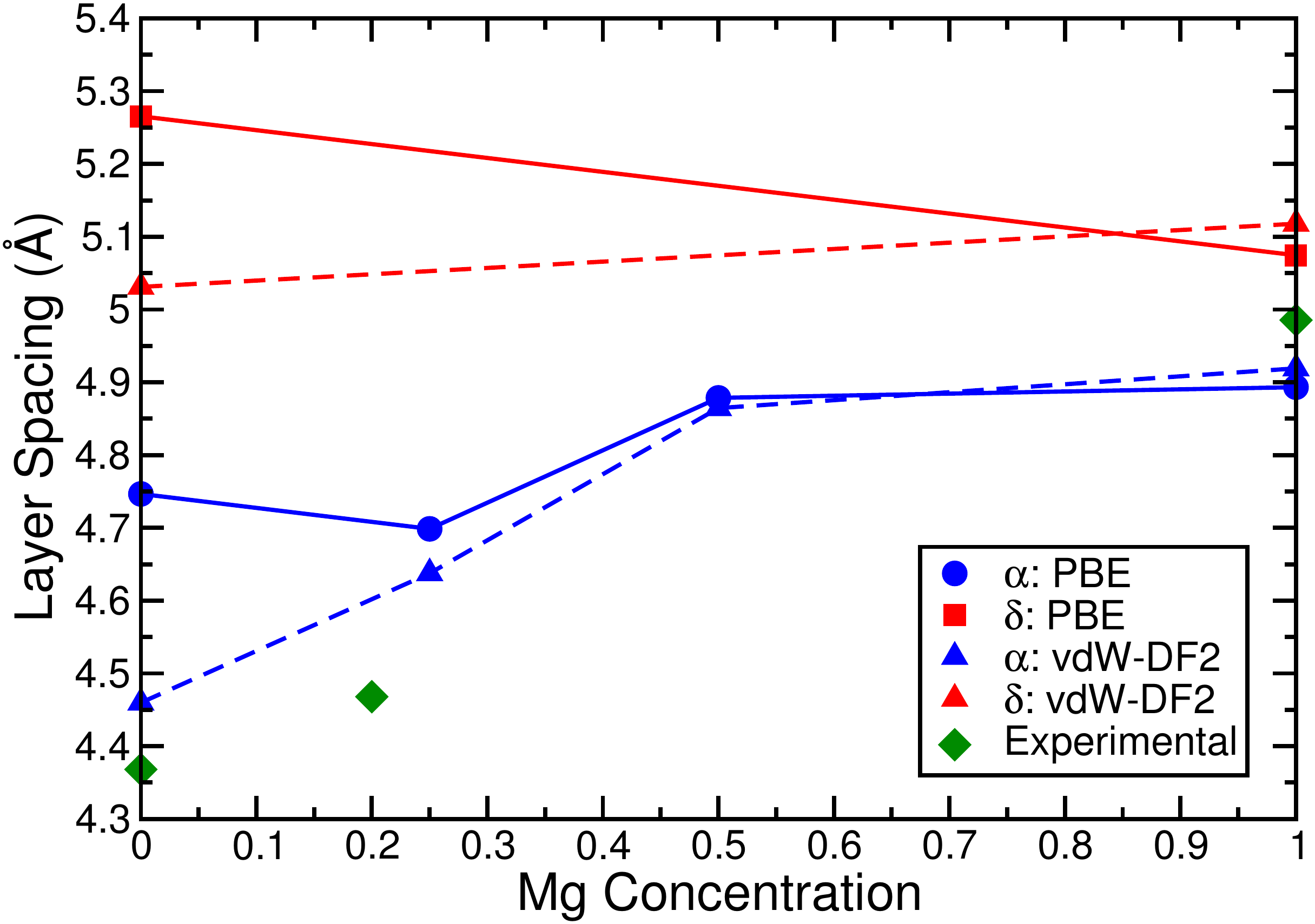}
\caption{
\label{fig:3} (Color online) 
Variation of layer spacing with Mg concentration in both $\alpha$ and $\delta$ phases. The experimental data points correspond to the pure $\alpha$-V$_2$O$_5$, intercalated Mg$_{0.2}$V$_2$O$_5$ and pure $\delta$-Mg$_1$V$_2$O$_5$.}
\end{figure}

Figure~\ref{fig:3} shows the variation of the V$_2$O$_5$ layer spacing (seen in Figure~\ref{fig:1}a and Figure~\ref{fig:1}b) as a function of Mg concentration in both the $\alpha$ (blue) and the $\delta$ (red) phases. In other layered materials, van der Waals interactions are known to cause layer binding in the deintercalated limit,\cite{Amatucci1996} which is not well described by standard DFT calculations.\cite{French2010,Rydberg2003} Therefore, in order to obtain a better estimate of the layer spacing values, additional calculations are performed using the vdW-DF2 functional,\cite{Lee2010,Klime2011} which includes the van der Waals interactions in addition to the Hubbard~$+$\emph{U}  Hamiltonian (for removing self-interaction errors).

The layer spacing values in Figure~\ref{fig:3} are taken from the relaxed ground states for $\alpha$ and $\delta$ in Figure~\ref{fig:2}a. The blue circles and red squares are obtained from PBE ($+$\emph{U}) calculations, while the blue and red triangles are calculated with vdW-DF2 ($+$\emph{U}). The experimental values listed (green diamonds) correspond to pure $\alpha$-V$_2$O$_5$,\cite{Enjalbert1986} Mg$_{0.2}$V$_2$O$_5$ reported by Pereira-Ramos \emph{et al.}\cite{Ramos1987} and pure $\delta$-Mg$_1$V$_2$O$_5$.\cite{Millet1998} As expected, the PBE and vdW-DF2 layer spacing values differ at complete demagnesiation ($\sim$ 0.3 $\textrm{\AA}$) but remain similar at all other Mg concentrations, where the layer spacing is determined by the electrostatics and short range repulsion. 

With increasing Mg concentration, the layer spacing increases significantly for $\alpha$-V$_2$O$_5$ ($\sim$~9\% increase from x$_\text{Mg} = 0$ to x$_\text{Mg} = 0.5$ while using vdW-DF2) but remains fairly constant in $\delta$-V$_2$O$_5$ ($\sim$~2\% increase from x$_\text{Mg} = 0$ to x$_\text{Mg} = 1$). However, the layer spacing in the $\delta$ phase remains higher than in the $\alpha$ phase across all Mg concentrations. Also, the layer spacing seen in the $\alpha$ phase (with vdW-DF2) benchmarks better with experimental layer spacing values at low Mg concentrations (up to x$_\text{Mg} = 0.2$) compared to the $\delta$ phase. Though including the van der Waals corrections in DFT leads to better agreement with the experimental V$_2$O$_5$ layer spacing, the Mg insertion voltage is overestimated\cite{Carrasco2014} (by ~18\% as compared to 6\% with PBE$+$\emph{U}), showing that PBE+\emph{U} describes the energetics more accurately than vdW-DF2. If the Mg$_x$V$_2$O$_5$ hull (Figure~\ref{fig:2}a) were to be calculated with vdW-DF2, we speculate that the energies of the demagnesiated structures will shift to higher values than PBE+\emph{U}, since van der Waals corrections tend to penalize under-binded (demagnesiated) structures.


\subsection{Mg diffusion barriers in V$_2$O$_5$}
\label{subsec:NEB}
\begin{figure}[h]
\includegraphics[width=\columnwidth]{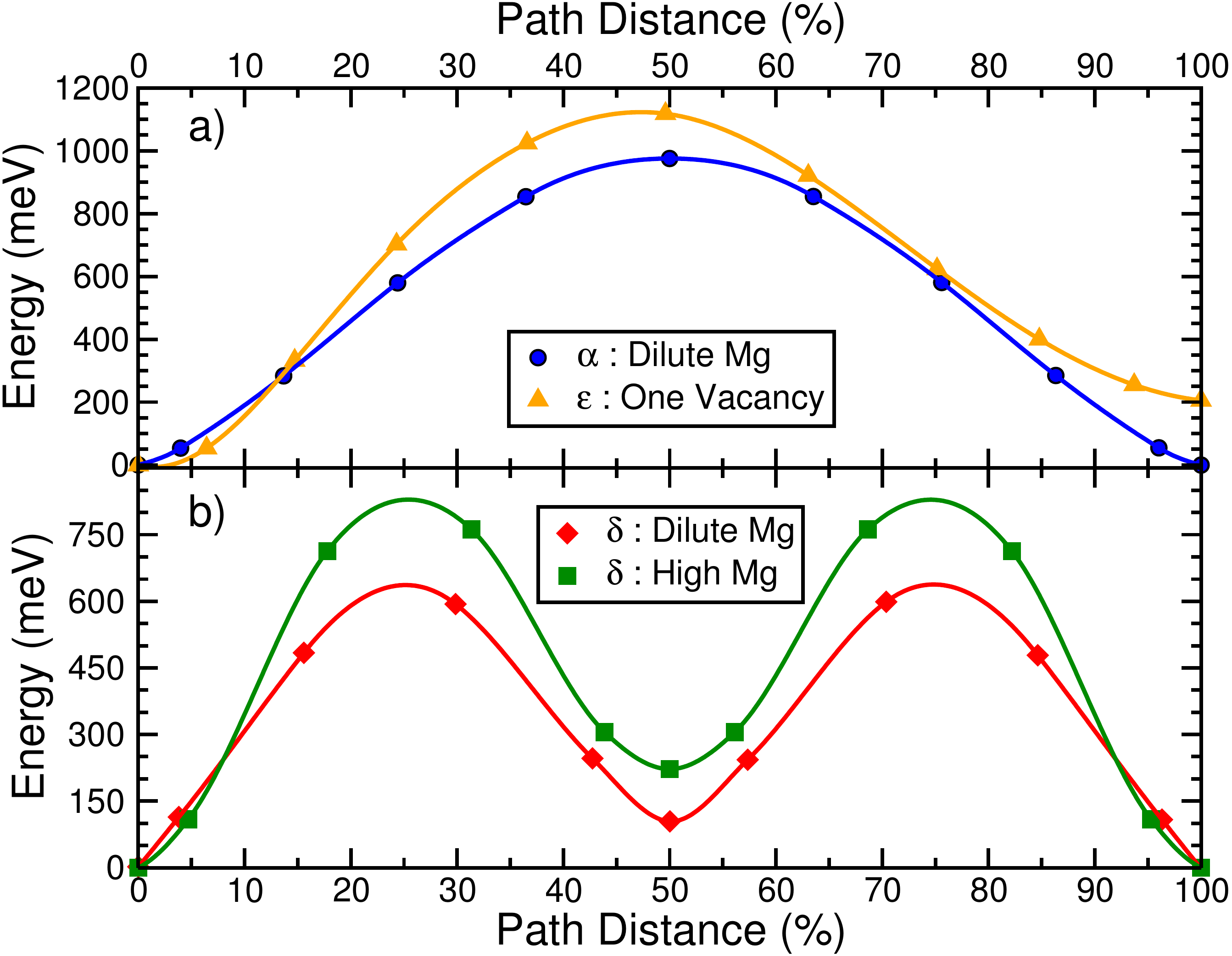}
\caption{
\label{fig:4} (Color online) 
(a) Activation barriers for Mg diffusion in select limiting cases in $\alpha$-V$_2$O$_5$ and (b) for Mg diffusion in $\delta$-V$_2$O$_5$ calculated through the NEB method.}
\end{figure}

To gain insight into the migration behavior of Mg in $\alpha$ and $\delta$ polymorphs, the calculated activation barriers using the NEB method are plotted in Figure~\ref{fig:4}. The migration energy is plotted along the diffusion path with the energies of the end points referenced to zero and the total path distance normalized to 100\%. The diffusion paths in both $\alpha$ and $\delta$ polymorphs correspond to the shortest Mg hop along the \emph{a}-direction as in Figure~\ref{fig:1}a and \ref{fig:1}b respectively and perpendicular to the \emph{b-c} plane in Figure~\ref{fig:1}c. The energy difference between the site with the highest energy along the path (the activated state) and the end points is the migration barrier. A simple random walk model for diffusion would predict that an increase in the activation barrier of $\sim$~60~meV would cause a drop in diffusivity by one order of magnitude at 298~K.

Specifically, we have performed four sets of calculations: dilute Mg concentration (x$_\text{Mg} = 0.08$) in the $\alpha$ phase (blue dots on Figure~\ref{fig:4}a), high Mg concentration (x$_\text{Mg} = 0.44$) in the $\alpha$ phase (orange triangles), dilute Mg concentration (x$_\text{Mg} = 0.08$) in the $\delta$ phase (red diamonds on Figure~\ref{fig:4}b) and high Mg concentrations (x$_\text{Mg} = 0.92$) in the $\delta$ phase (green squares). Due to the mechanical instability of the $\alpha$ phase at high Mg concentrations, we performed NEB calculations in the $\epsilon$ phase. Because the $\epsilon$ phase has a specific Mg ordering, migration to an equivalent site requires two symmetrically equivalent hops. The path in the orange triangles of Figure~\ref{fig:4}a therefore only shows one half of the total path.

The data in Figure~\ref{fig:4} illustrates that the barriers in the $\delta$ phase ($\sim$~600~$-$~760~meV) are consistently much lower than in the $\alpha$-phase ($\sim$~975~$-$~1120~meV), with the respective migration energies adopting ``valley" and ``plateau" shapes. Upon addition of Mg the migration barriers in $\alpha$ and $\delta$ both increase. The differences in the magnitude of the migration barriers and the shape of the migration energies between the $\alpha$ and $\delta$ can be explained by considering the changes in the coordination environment of Mg along the diffusion path. For example, in the $\alpha$ phase, Mg migrates between adjacent 8-fold coordinated sites through a shared 3-fold coordinated site (activated state), a net 8$\rightarrow$3$\rightarrow$8 coordination change, while in the $\delta$ phase Mg migrates between adjacent 6-fold coordinated sites through two 3-fold coordinated sites separated by a metastable 5-fold coordinated ``valley", a net 6$\rightarrow$3$\rightarrow$5$\rightarrow$3$\rightarrow$6 coordination change. Hence, the lower barriers of the $\delta$ phase compared to the $\alpha$ phase are likely due to the smaller coordination changes and the higher layer spacing in $\delta$ than $\alpha$ as seen in Figure~\ref{fig:3}. The indication of superior diffusivity of Mg in $\delta$-V$_2$O$_5$ motivates investigating the intercalation properties of Mg in the $\delta$ phase further.

\subsection{Cluster expansion on Mg in $\delta$-V$_2$O$_5$ and temperature-composition phase diagram}
\label{subsec:CE}
\begin{figure}[h]
\includegraphics[width=\columnwidth]{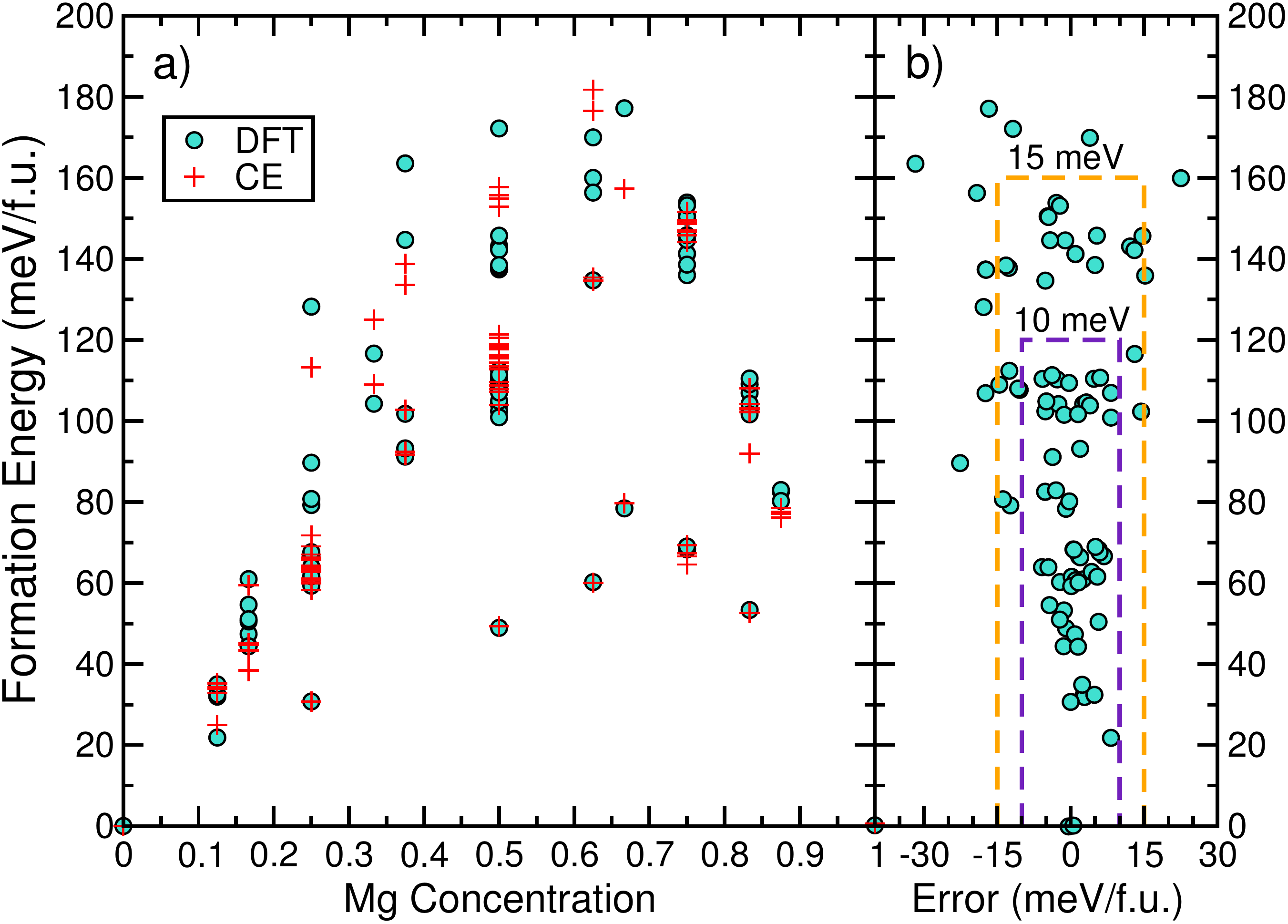}
\caption{
\label{fig:5} (Color online) 
(a) DFT and Cluster expansion predicted formation energies are plotted on the vertical scale with respect to different Mg concentrations on the horizontal scale. (b) The staircase plot indicates the errors in energies encountered for structures using the cluster expansion (horizontal scale) with respect to their respective distances from the hull (vertical scale).}
\end{figure}

Consistent with the data in Figure~\ref{fig:2}a all Mg-Va arrangements have higher energy than the linear combination of $\delta$-V$_2$O$_5$ and $\delta$-MgV$_2$O$_5$, supporting phase separation on the $\delta$ lattice as illustrated in Figure~\ref{fig:5}a, where the zero on the energy scale is referenced to the DFT calculated end members of the $\delta$ phase. A total of 97 Mg-Va configurations, across Mg concentrations are used to construct the CE, which encompasses 13 clusters with a RMSE of $\sim$~9~meV/f.u. The CEs Weighted Cross Validation (WCV) score of $\sim$~12.25~meV/f.u. indicates a very good match with the current input set and good predictive capability. In Figure~\ref{fig:5}b the staircase plot displays the error in predicting the formation energies of different Mg-Va configurations by the CE against their respective DFT formation energies. A good CE will have lower errors for configurations that are closer to the hull, i.e. shorter absolute distance from the ground state hull, and higher errors for configurations that are further away from the hull. The current CE displays errors below 10~meV/f.u. for most structures whose formation energies are smaller than 120~meV/f.u. Also, it can be seen in Figure~\ref{fig:5}b that the structures with the highest errors in the formation energy prediction normally have formation energies greater than 125~meV/f.u.

\begin{figure}[h]
\includegraphics[width=\columnwidth]{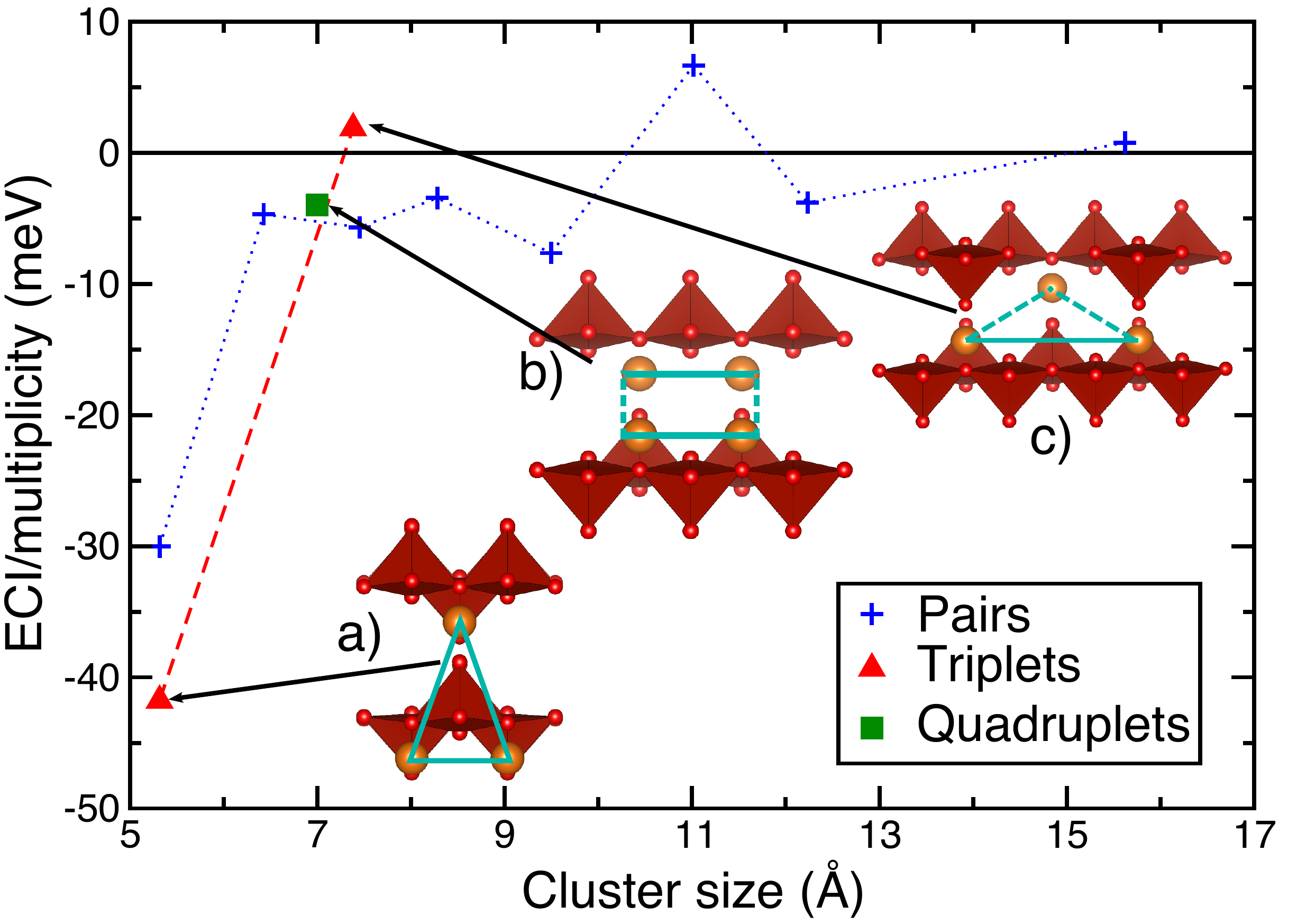}
\caption{
\label{fig:6} (Color online) 
ECI of the clusters vs. their respective cluster size are plotted. The insets (a) and (c) display the triplet terms and inset (b) shows the quadruplet term with the solid blue lines indicating in-plane interactions and the dotted blue lines indicating out-of-plane interactions. All insets are displayed on the \emph{a-b} plane.}
\end{figure}

The ECIs for the clusters in the CE, normalized by their multiplicity and plotted against their respective cluster sizes, are displayed in Figure~\ref{fig:6}. The size of a given cluster is indicated by its longest dimension; for example, in a triplet the cluster size is given by its longest pair. Negative pair terms indicate `attraction' (i.e.\ Mg-Mg and Va-Va pairs are favored) and positive pair terms indicate `repulsion' (i.e.\ Mg-Va pairs are favored). The figures inside the graph show the triplets and the quadruplet used in the current CE with the solid lines indicating interactions in the \emph{a-b} plane and dotted lines indicating interactions out of plane (\emph{b} is the direction perpendicular to the V$_2$O$_5$ layers). The orange circles indicate Mg atoms. The data in Figure~\ref{fig:6} illustrates that the most dominant (highest absolute ECI value) cluster of the CE is a triplet where Mg ions are along the \emph{a-b} plane (as shown in Figure~\ref{fig:1}b). The most dominant pair term is attractive and is the longest pair of the most dominant triplet. The negative sign of the dominant triplet and the dominant pair terms implies that there are 2 possible configurations containing Mg which are stabilized: \emph{i})~all three sites are occupied by Mg, and \emph{ii})~only one of the three sites is occupied by Mg, consistent with the sign convention adopted in the CE ($\sigma_i = 1$ for occupied Mg site and $\sigma_i = -1$ for a vacancy). 

\begin{figure}[h]
\includegraphics[width=\columnwidth]{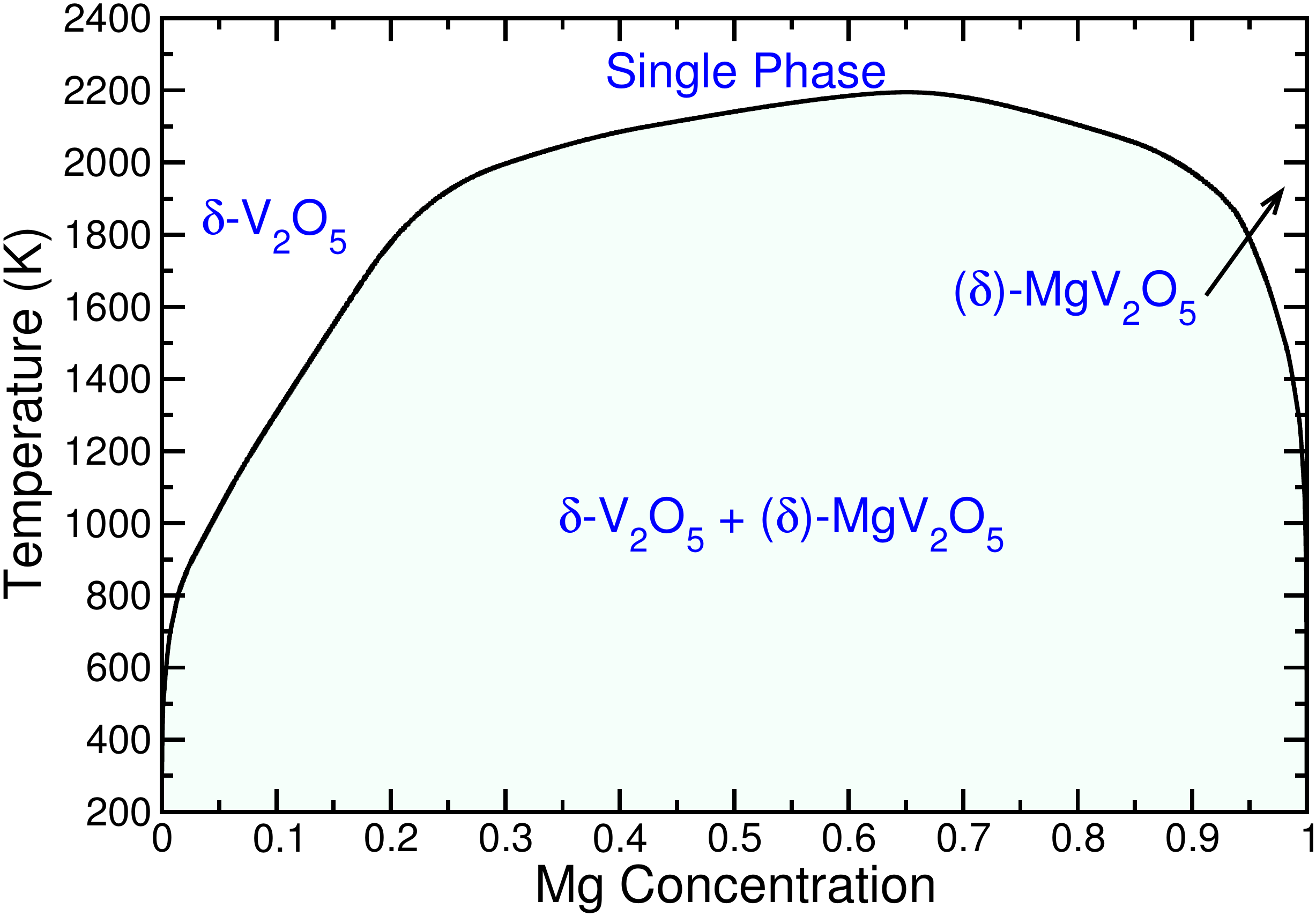}
\caption{
\label{fig:7} (Color online) 
Mg-V$_2$O$_5$ intercalation phase diagram for the $\delta$ phase. The black line indicates the phase boundary between the single and two phase regions obtained from Monte Carlo simulations of the CE.}
\end{figure}

The temperature-concentration phase diagram for Mg intercalation into $\delta$-V$_2$O$_5$ is displayed in Figure~\ref{fig:7}. The black line traces the phase boundary between the single and two phase regions, obtained from Monte Carlo simulations with the numerical hysteresis removed by free energy integration. Consistent with the $\delta$ hull in Figure~\ref{fig:2}a, the Mg-V$_2$O$_5$ is a phase separating system at room temperature with extremely low solubilities at either ends ($<1$\%). Note that only the solid $\delta$-phase is considered in this phase diagram. In reality, the high temperature part of the phase diagram would probably form a eutectic since pure V$_2$O$_5$ melts at $\sim$~954~K.\cite{Haynes}

\section{Discussion}
\label{sec:discussion}
In this work, we have performed a first-principles investigation of Mg intercalation into orthorhombic V$_2$O$_5$. Specifically, we investigated the $\alpha$ and $\delta$ polymorphs using DFT calculations, evaluating their respective ground state hulls, subsequent voltage curves, and their Mg migration barriers. For the $\delta$ polymorph, we constructed the composition-temperature phase diagram using the CE and GMC approach. The theoretical data we have collected sheds light not only on the existing experiments intercalating Mg into V$_2$O$_5$, but also provides a practical strategy to improve performance.

From a thorough comparison of the experimental data available in the literature to the calculations performed in this work, we conclude that by synthesizing V$_2$O$_5$ and intercalating Mg (i.e. beginning in the charged state), the structure remains in the $\alpha$ phase. For example, in the experimental voltage curves\cite{Amatucci2001a,Ramos1987,Gershinsky2013,Novak1993,Yu2004} the characteristic plateau followed by a drop at x$_\text{Mg} \sim 0.5$ compares well with the computed voltage curve for the $\alpha$ phase (Figure~\ref{fig:2}b) which shows a similar voltage drop corresponding to the $\epsilon$ ordering while $\delta$-V$_2$O$_5$ would show no such drop. In X-ray diffraction (XRD) data in the literature on magnesiated V$_2$O$_5$, no additional peaks which would indicate the formation of the $\delta$ phase have been observed.\cite{Ramos1987,Gershinsky2013,Novak1994} Also, the observed increase in the layer spacing\cite{Gershinsky2013} is consistent with the computed predictions of layer expansion in the $\alpha$ phase until x$_\text{Mg} = 0.5$ (Figure~\ref{fig:3}) rather than the $\delta$ phase which has a minimal increase in layer spacing from x$_\text{Mg} = 0$ to x$_\text{Mg} = 1$. The migration barriers for Mg in the $\alpha$ phase are high ($\sim$~975~meV as seen in Figure~\ref{fig:4}a), and indeed, reversible Mg insertion can be reliably achieved only when the diffusion length is greatly reduced (i.e. in thin films and nano-powders) and at very low rates (i.e.\ $\sim$~0.5~$\mu$A/cm$^2$ by Gershinsky \emph{et al.}\cite{Gershinsky2013}). Magnesiation past the $\epsilon$-phase (x$_\text{Mg} \sim 0.5$) is expected to be difficult as the potential drops thereby reducing the driving force for Mg insertion, and the Mg migration barrier increases with Mg concentration in $\alpha$ (Figure~\ref{fig:4}a). While the driving force to transform from $\alpha \rightarrow \delta$ is small up to x$_\text{Mg} \sim 0.5$ (as in Figure~\ref{fig:2}a), it steeply increases thereafter, leading us to speculate that further magnesiation would lead to the formation of a fully magnesiated $\delta$-MgV$_2$O$_5$ on the surface.

\begin{figure}[h]
\includegraphics[width=\columnwidth]{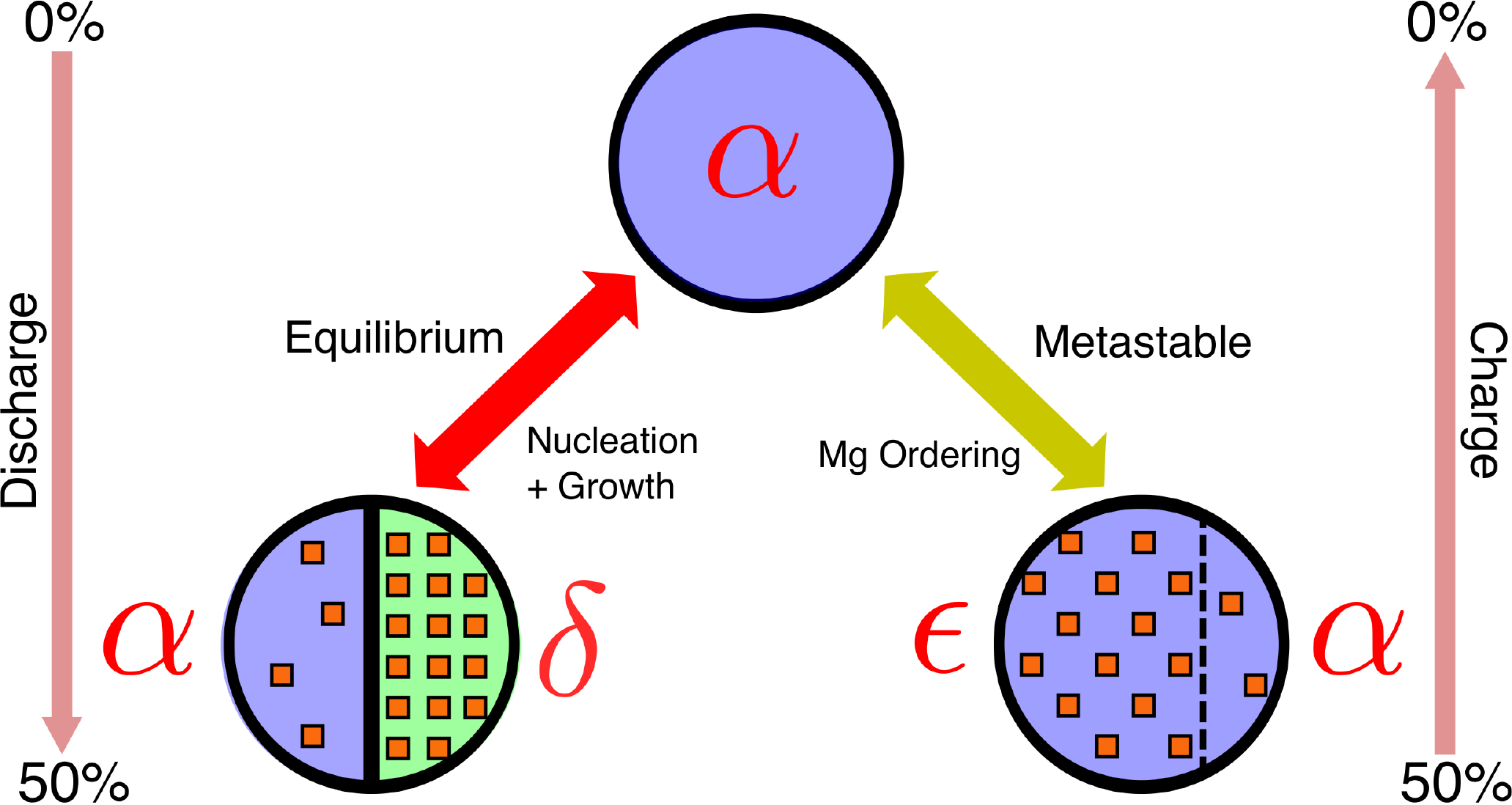}
\caption{
\label{fig:8} (Color online) 
Possible intercalation pathways for Mg in V$_2$O$_5$ up to x$_\text{Mg} = 0.5$. The left half corresponds to the equilibrium case where the $\delta$ phase nucleates and grows in a supersaturated $\alpha$ phase, with a well-defined interface between the two phases and the right half corresponds to the Mg atoms ordering into the metastable $\epsilon$ phase and the lack of a well defined interface in this case since $\epsilon$ and $\alpha$ have the same V$_2$O$_5$ layer stacking.}
\end{figure}

Our thinking on the magnesiation process of V$_2$O$_5$ is summarized in Figure~\ref{fig:8}. The ground state hull in Figure~\ref{fig:2}a, suggests that under equilibrium conditions the Mg insertion mechanism is through a two-phase reaction, by nucleation and growth of magnesiated $\delta$ phase from supersaturated $\alpha$, rather than through the metastable formation of the $\epsilon$ phase. These two reaction pathways (cycling between 0 and 50\% state of charge) are illustrated schematically in Figure~\ref{fig:8}, with the orange squares representing Mg atoms. If nucleation and growth of the fully magnesiated $\delta$ phase (i.e.\ x$_\text{Mg} = 1$) were to occur, there would be no inherent upper limit to magnesium insertion up to x$_\text{Mg} \sim 1$. However, the metastable insertion path of Mg in the $\alpha$ phase, which once fully converted to $\epsilon$ phase remains at x$_\text{Mg} \sim 0.5$, is more consistent with experiments. The reason the system follows the metastable insertion path through $\alpha$ is that the equilibrium path ($\alpha$-V$_2$O$_5$ to $\delta$-MgV$_2$O$_5$), requires structural rearrangement of the host structure through the translation of V$_2$O$_5$ layers, which may kinetically be difficult once some Mg is inserted and more strongly bonds the layers. Also, a nucleation-growth process involves high interfacial energies and may lead to low rates. A similar metastable solid solution transformation has been predicted and documented for other thermodynamic phase separating systems.\cite{Malik2011,Kang2013,Ganapathy2014}

While our calculations, supported by experimental data, suggest that the host V$_2$O$_5$ structure remains in the $\alpha$ phase upon Mg intercalation, they also suggest an approach to substantially improve the electrochemical properties by cycling Mg beginning in the $\delta$ phase. Mg in $\delta$-V$_2$O$_5$ not only possesses a higher average voltage compared to $\alpha$ ($\sim$~120~mV higher as seen in Figure~\ref{fig:2}b), but also a significantly better mobility ($\sim$~600~$-$~760~meV compared to $\sim 975 - 1120$~meV) which accounts for approximately 5 orders of magnitude improvement in the diffusivity at room temperature (Figure~\ref{fig:4}). Prior computations have reported higher migration barriers in the $\delta$ phase compared to the $\alpha$ phase in the charged limit, in contrast to our calculations in Figure~\ref{fig:4},\cite{Zhou2014} which we attribute to the authors allowing only Mg and nearby oxygen ions to relax in their NEB calculations. In order to cycle Mg in the $\delta$ phase, V$_2$O$_5$ must be prepared in the fully discharged state ($\delta$-Mg$_1$V$_2$O$_5$), where the $\delta$ phase is thermodynamically stable. Fortunately, the synthesis of $\delta$-MgV$_2$O$_5$ is well established in the literature.\cite{Millet1998}

Since at intermediate Mg concentrations the equilibrium state is a coexistence between the demagnesiated $\alpha$-phase and the fully magnesiated $\delta$-phase, the $\delta$ phase must remain metastable over a wide Mg concentration range to ensure higher capacities. If the $\delta$-phase is not metastable, transformation to the $\alpha$-phase will take place. We speculate that the possibility of $\delta$ phase metastability is likely, given that nucleation and growth of the $\alpha$ phase requires restructuring of the host lattice, and the absence of mechanically unstable Mg configurations (even at x$_\text{Mg} = 0$) in $\delta$ (Figure~\ref{fig:2}a) in our calculations. Also, an applied (over)underpotential is required to access a metastable (de)insertion path, which can be quantified by the difference between the metastable and equilibrium voltage curves in Figure~\ref{fig:2}b. For example, to avoid the equilibrium path, an applied underpotential of $\sim$~800~mV is required to insert Mg and retain the $\alpha$-V$_2$O$_5$ structure, but only $\sim$~400~mV is required to remove Mg and retain the $\delta$-MgV$_2$O$_5$ structure, which supports the possibility of a metastable $\delta$ phase. 

Assuming the $\delta$-MgV$_2$O$_5$ phase remains metastable, the temperature-composition phase diagram computed for Mg in $\delta$-V$_2$O$_5$ using the CE (Figure~\ref{fig:7}) indicates a phase separating behavior with negligible solubility at both end members at room temperature. By investigating the dominant interactions (ECIs) that contribute to the CE, we gain some insight into the possible intercalation mechanism. The dominant Mg-Va interactions, specifically the triplet and the nearest interlayer pair as seen in Figure~\ref{fig:6}, are entirely contained in the \emph{a-b} plane, which indicates that the $\delta$-V$_2$O$_5$ host lattice will contain fully magnesiated and fully demagnesiated domains separated by an interface along an \emph{a-b} plane. Hence, Mg insertion into the 3D $\delta$-V$_2$O$_5$ structure can be effectively described by considering the interactions in each 2D \emph{a-b} plane.

\begin{figure}[h]
\includegraphics[width=\columnwidth]{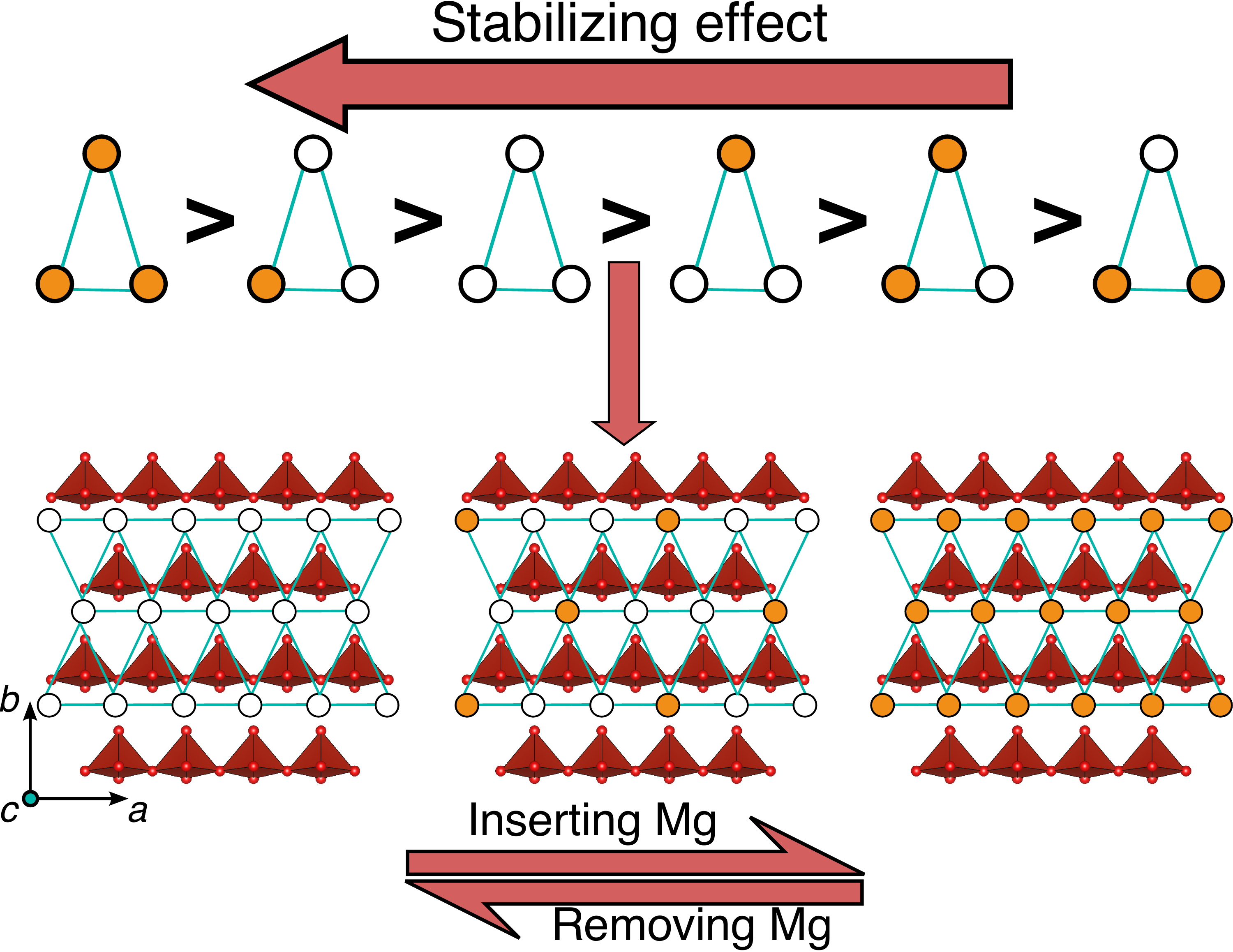}
\caption{
\label{fig:9} (Color online) 
Interplay between the dominant pair and triplet terms of the CE stabilizing different Mg-Va arrangements.}
\end{figure}

Figure~\ref{fig:9} illustrates the interplay between these dominant pair and triplet terms which results in the specific sequence of Mg-Va configurations in terms of their relative stability. The orange circles indicate Mg atoms, the hollow circles the vacancies, and all insets are viewed in the \emph{a-b} plane. Given the sign convention used in the CE ($\sigma_i = 1$ for Mg and $\sigma_i = -1$ for Va) and the negative sign of the dominant pair and triplet, the formation of Mg-Mg and Va-Va pairs are favored while triplets containing one or three Mg atoms are favored. Thus, a fully occupied triplet is most stable due to favorable contributions from both the triplet ($\sim$~$-$40~meV) and the two longest pair terms ($\sim$~$-$60~meV in total) resulting in a net stabilization of $\sim$~$-$100~meV, while the triplet with two Mg atoms forming the shortest pair and a vacancy at the apex is least stable due to unfavorable contributions from both the pairs and the triplet resulting in a destabilizing contribution of $\sim$~$+$100~meV. 

The bottom half of Figure~\ref{fig:9} illustrates a sample sequence in which Mg atoms fill up sites on a given \emph{a-b} plane. The fully magnesiated structure (right inset) is highly stabilized due to the presence of fully filled triplets ($\sim$~$-$100~meV/triplet) while the fully demagnesiated structure (right inset) is stabilized to a lesser extent ($\sim$~$-$20~meV/triplet). At an intermediate composition, the Mg atoms will arrange themselves in such a way that the number of fully filled and one-third filled triplets ($\sim$~$-$40~meV/triplet, depicted in the centre inset) is maximized. Since one-third filled triplets stabilize a structure more than triplets containing two Mg atoms, non-phase separated configurations at low Mg concentrations (x$_\text{Mg} < 0.33$) will be more stabilized than those at high Mg concentrations (x$_\text{Mg} > 0.66$), as indicated by the higher solubilities at lower Mg concentrations in the phase diagram shown in Figure~\ref{fig:7} at high temperatures. 

Since the occurrence of fully magnesiated and demagnesiated \emph{a-b} planes is highly stabilized, the intercalation of Mg in the 3D $\delta$-V$_2$O$_5$ structure will then progress via propagation of fully magnesiated \emph{a-b} planes along the \emph{c}-axis. With additional applied overpotential, not only can the $\delta$ phase be retained, but also a non-equilibrium solid solution intercalation pathway in $\delta$ can be thermodynamically accessible, leading to further improved kinetics.\cite{Malik2011} An estimate for the additional overpotential required can be computed by considering the lowest energy structure at x$_\text{Mg} = 0.83$ in Figure~\ref{fig:5}a, whose formation energy is 53~meV/Mg, resulting in an approximate additional overpotential requirement of $\sim$~320~mV. Therefore, the net overpotential required to access a solid-solution transformation path entirely in the $\delta$ phase upon charge is $\sim$~720~mV, which is comparable to the underpotential applied ($\sim$~800~mV) to remain in the metastable $\alpha$ phase upon discharge. Hence, we suggest that the electrochemical performance of Mg in V$_2$O$_5$ can be improved by beginning cycling in the discharged state, $\delta$-MgV$_2$O$_5$, with the prospect of improved voltage, capacity, and kinetics. 

\section{Conclusions}
\label{sec:conclusion}
In this work, we have used first-principles calculations to perform an in-depth investigation of Mg intercalation in the orthorhombic $\alpha$ and $\delta$ polymorphs of V$_2$O$_5$ to evaluate their suitability as high energy density cathode materials for Mg-ion batteries. Specifically, we computed the ground state hulls and the activation energies for Mg migration in both polymorphs. For the $\delta$ polymorph we calculated the temperature-composition phase diagram. The equilibrium state of Mg$_x$V$_2$O$_5$ ($0 < $ x$_\text{Mg} < 1$) is determined to be a two-phase coexistence between the fully magnesiated $\delta$-MgV$_2$O$_5$ and fully demagnesiated $\alpha$-V$_2$O$_5$ phases. NEB calculations indicate that room-temperature Mg migration is several orders of magnitude faster in the $\delta$ phase (E$_\text{m} \sim 600-760$~meV) than in the $\alpha$ phase (E$_\text{m} \sim 975-1120$~meV). 

By comparing the calculated voltage curves and changes in the layer spacing with intercalation with available experimental data on Mg insertion in V$_2$O$_5$, we conclude that the $\alpha$ phase likely remains metastable when Mg is initially inserted into fully demagnesiated $\alpha$-V$_2$O$_5$. Although the computed $\alpha$ phase migration barriers indicate poor Mg mobility, consistent with reversible Mg intercalation being achievable exclusively at very low rates and in small particles, $\alpha$-V$_2$O$_5$ is still one of only three known cathode materials where reversible cycling of Mg is possible at all (along with chevrel Mo$_6$S$_8$ and layered MoO$_3$). 

Therefore, our finding that the $\delta$-V$_2$O$_5$ polymorph displays vastly superior Mg mobility as well as a modest increase in voltage compared to the $\alpha$ phase is especially promising, assuming that the $\delta$-V$_2$O$_5$ host structure can remain stable or metastable across a wide Mg concentration range. Fortunately, the $\delta$ polymorph is thermodynamically stable in the fully discharged state and its synthesis procedure well known. 

From our first-principles calculations of the formation energies of several Mg orderings in the $\delta$-V$_2$O$_5$ host structure and the resulting computed temperature-composition phase diagram, we have also gained insight into the possible mechanism of Mg intercalation within the $\delta$ host structure. At room temperature, Mg displays strong phase-separating behavior with negligible solid-solution in the end-member phases and favors the formation of either completely full or empty \emph{a-b} planes, which are perpendicular to the layers formed by the connecting VO$_5$ pyramids, suggesting an intercalation mechanism based on nucleation and growth through the propagation of an \emph{a-b} interface along the \emph{c}-axis.

\begin{acknowledgements}
The current work is fully supported by the Joint Center for Energy Storage Research (JCESR), an Energy Innovation Hub funded by the U.S. Department of Energy, Office of Science and Basic Energy Sciences. This study was supported by Subcontract 3F-31144. The authors thank the National Energy Research Scientific Computing Center (NERSC) for providing computing resources. S.G.\ would like to thank William Richards at MIT for fruitful feedback and suggestions.
\end{acknowledgements}


\bibliography{library}

\end{document}